\documentclass{article}
\usepackage{spconf,graphicx}
\usepackage{booktabs}
\usepackage{url}
\usepackage{multirow}
\usepackage{makecell}
\usepackage{booktabs}
\usepackage{nicematrix}
\usepackage{stackrel}
\usepackage{tabularx}
\usepackage{xcolor}
\usepackage{soul}
\usepackage{adjustbox}
\usepackage{float}
\usepackage{subfig}
\usepackage{xcolor,colortbl}
\usepackage{pifont}

\definecolor{LightCyanBG}{rgb}{0.85,0.92,0.97}
\definecolor{lightcornflowerblue}{rgb}{0.6, 0.81, 0.93}
\definecolor{cambridgeblue}{rgb}{0.64, 0.76, 0.68}
\definecolor{navajowhite}{rgb}{0.85, 0.85, 0.85}
\newcolumntype{a}{>{\columncolor{lightcornflowerblue}}c}
\newcolumntype{b}{>{\columncolor{cambridgeblue}}c}
\newcolumntype{d}{>{\columncolor{navajowhite}}c}
\usepackage{hhline}
\newcommand{\cmark}{\ding{51}}%
\newcommand{\xmark}{\ding{55}}%
\usepackage{amssymb,amsmath,amsthm,enumitem}

\usepackage{colortbl}

\usepackage[pagebackref,breaklinks,colorlinks]{hyperref}

\usepackage{enumitem}
\setlist{nosep, leftmargin=14pt}

\usepackage{mwe} 

\title{SAM3D: Segment Anything Model in Volumetric Medical Images}
\name{Nhat-Tan Bui$^{1*}$,  Dinh-Hieu Hoang$^{2,3*}$, Minh-Triet Tran$^{2,3}$, Gianfranco Doretto$^{4}$} 
{Donald Adjeroh$^{4}$, Brijesh Patel$^{4}$, Arabinda Choudhary$^{5}$, Ngan Le$^{1}$}
\address{$^{1}$AICV Lab, University of Arkansas, Arkansas, USA \\
$^{2}$University of Science, Vietnam National University, Ho Chi Minh City, Vietnam \\
$^{3}$John von Neumann Institute, Vietnam National University, Ho Chi Minh City, Vietnam\\
$^{4}$West Virginia University, West Virginia, USA\\
$^{5}$University of Arkansas for Medical Sciences, Arkansas, USA \\
}

\begin{document}
\maketitle
\def\thefootnote{*}\footnotetext{Equal contribution}\def\thefootnote{\arabic{footnote}}
\begin{abstract}
Image segmentation remains a pivotal component in medical image analysis, aiding in the extraction of critical information for precise diagnostic practices. With the advent of deep learning, automated image segmentation methods have risen to prominence, showcasing exceptional proficiency in processing medical imagery. Motivated by the Segment Anything Model (SAM)—a foundational model renowned for its remarkable precision and robust generalization capabilities in segmenting 2D natural images—we introduce SAM3D, an innovative adaptation tailored for 3D volumetric medical image analysis. Unlike current SAM-based methods that segment volumetric data by converting the volume into separate 2D slices for individual analysis, our SAM3D model processes the entire 3D volume image in a unified approach. Extensive experiments are conducted on multiple medical image datasets to demonstrate that our network attains competitive results compared with other state-of-the-art methods in 3D medical segmentation tasks while being significantly efficient in terms of parameters. Code and checkpoints are available at \url{https://github.com/UARK-AICV/SAM3D}.
\end{abstract}
\vspace{-1em}

\section{Introduction}
\vspace{-0.5em}
Volumetric segmentation is crucial in medical image analysis, finding applications in pathology diagnosis, surgical planning, and computer-aided diagnosis. Volumetric medical images like CT, MRI, OCT, and DBT offer a 3D view of anatomical structures. Segmentation identifies regions of interest for better interpretation.

Deep learning, particularly UNet \cite{unet} and variants \cite{nnunet, transunet, swinunet}, made strides in 3D medical segmentation but faced limitations. Transformer-based models like Vision Transformer (ViT) \cite{vit2020} and Swin-UNet \cite{swinunet} showed promise in capturing long-range relationships. Combining CNNs and Transformers in models like TransUNet \cite{transunet}, UNETR \cite{unetr}, and HiFormer \cite{hiformer}, yielded promising results. However, these models prioritize precision, leading to increased complexity and training time. Leveraging pretrained models offers an alternative. SAM, a transformer-based model pretrained on large-scale datasets, has shown generalizability in segmentation tasks. SAM-based models for medical images piqued interest.

This work introduces \textbf{SAM3D}, an architecture for volumetric medical segmentation, combining the SAM encoder and a lightweight 3D CNN decoder. Unlike traditional slice-by-slice processing, SAM3D extracts features across the entire volume, improving segmentation while maintaining simplicity and computational efficiency. Contributions include applying the SAM encoder to process 3D volumes, designing SAM3D for effective 3D medical segmentation, and validating its performance on various datasets, such as ACDC \cite{acdc}, Synapse \cite{synapse}, MSD BraTS \cite{lungandbrats}, and MSD Lung \cite{lungandbrats}. SAM3D demonstrates competitive results, marking a novel approach to 3D volumetric imaging.
\vspace{-1em}
\begin{figure*}[!t]
\begin{center}
\includegraphics[width=0.85\textwidth]{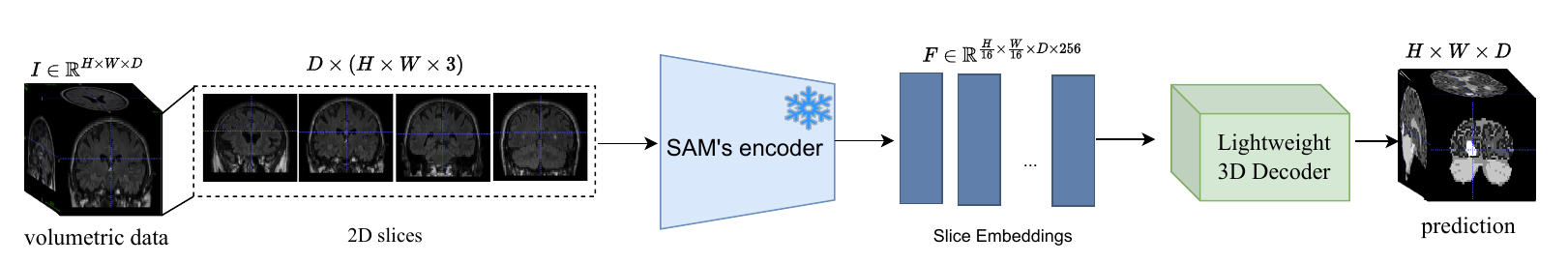}
\vspace{-1em}
\caption{Overall architecture of the proposed SAM3D. Given a volumetric image $I \in \mathbb{R}^{H\times W\times D}$, SAM3D initially applies SAM to process each of the $D$ slices individually, producing slice embeddings denoted as $F \in \mathbb{R}^{\frac{H}{16}\times \frac{W}{16}\times D\times 256}$. These embeddings are then decoded by a lightweight 3D decoder, ultimately yielding the segmentation prediction.} 
\vspace{-2em}
\label{fig:SAM3D}
\end{center}
\end{figure*}

\section{Related Works}
\vspace{-0.5em}
\textbf{Segmentation Methods using CNNs and Transformer:} Various methods leverage a combination of CNNs and Transformer architectures for segmentation tasks. TransUNet \cite{transunet} integrates CNNs and Transformer within a U-shaped architecture to capture both local and global information. Swin-Unet \cite{swinunet} replaces U-Net's convolutional blocks with Swin Transformer blocks. nnUNet \cite{nnunet} introduces a self-adapting framework for 2D and 3D medical segmentation. MISSFormer \cite{missformer} enhances hierarchical feature representation using Enhanced Transformer Blocks. TransDeepLab \cite{transdeeplab} combines Swin Transformer blocks with ASPP module and cross-contextual attention. HiFormer \cite{hiformer} introduces the Double-Level Fusion (DLF) module. UNETR \cite{unetr} encodes 3D input patches with Transformers and combines feature extraction with a CNNs-based decoder. Swin UNETR \cite{swinunetr} enhances UNETR with Swin Transformer blocks. nnFormer \cite{nnformer} interleaves CNNs and Transformer blocks with feature pyramids. UNETR++ \cite{unetr++} introduces the efficient paired-attention (EPA) module.

\noindent \textbf{Segment Anything Model (SAM) in medical:} SAM \cite{sam} is a foundational model for natural image segmentation that can be guided by prompts. It comprises an image encoder, prompt encoder, and lightweight mask decoder, trained on promptable segmentation tasks. SAMed \cite{samed} adapts SAM to medical images using a series of finetuning strategies. MedSAM \cite{medsam} re-trains SAM on a union of medical image datasets.

In contrast to existing approaches that involve fine-tuning SAM and handling 3D images as sets of 2D slices, and unlike the conventional CNNs/Transformer-based methods that typically require large model designs, our proposed SAM3D effectively and efficiently harnesses SAM's capabilities for 3D medical segmentation. It does so without the need for large model architectures or depending on slice-by-slice predictions. This approach enhances the model's ability to perceive anatomical structures and capture global information.
\vspace{-1em}

\section{Method}
\label{sec:Method}
\vspace{-0.5em}
In this section, we introduce our model, SAM3D, and explain the rationale behind its simple design. Our goal is to leverage SAM without the need for extensive parameter retraining or complex task-specific modules.

\noindent
\textbf{Overall Architecture. }
SAM was trained on an extensive dataset comprising 1 million images and 1.1 billion masks, and it features a robust image encoder tailored for natural images. However, applying SAM directly to 3D medical images poses challenges due to inherent domain differences. We posit that the SAM image encoder retains valuable low-level features, e.g. edges and boundaries, which have relevance across various image domains.


In contrast to SAMed \cite{samed} and MedSAM \cite{medsam}, where all three components of SAM are fine-tuned, our approach involves freezing SAM's image encoder and training a new lightweight 3D decoder. SAM3D leverages SAM by initially processing images slice by slice and then incorporating a lightweight 3D decoder to capture depth-wise relationships between slices. The overall architecture of SAM3D is depicted in Figure \ref{fig:SAM3D} and can be summarized as follows: a volumetric input $I \in \mathbb{R}^{H\times W\times D}$ is divided into $D$ 2D slices, each of dimension $H\times W$. We duplicate each channel three times to generate the slices that have dimension of $H \times W \times 3$. The pretrained SAM encoder processes these slices, generating 3D slice embeddings denoted as $F$. The depth-wise relationships among these slice embeddings are effectively captured by our proposed 3D decoder. Additionally, we remove the prompt encoder from SAM to ensure that feature extraction remains uninhibited across different modalities.

\noindent
\textbf{Encoder. }
SAM's image encoder extracts robust low-level information. Thus, it is plausible to tackle the notorious weak boundary in the medical image domain by using features extracted by SAM's image encoder. Formally, let $I\in\mathbb{R}^{H\times W\times D}$ be the input, and $Enc$ represent the slice encoder. We split $I$ into $D$ slices $I_i$ along the depth dimension, each slice is in  $3\times H\times W$, and feed them into $Enc$. The output slice embeddings are stacked and transposed to obtain the final 3D slice embeddings $F = [f_i]_{i=1}^D$.
\begin{equation}
    f_i=Enc(I_i), \text{where }f_i\in\mathbb{R}^{\frac{H}{16}\times\frac{W}{16}\times 256}
\end{equation}
We stack these slice embeddings and transpose the result to obtain the final 3D slice embedding, $F = [f_i]_{i=1}^D, F\in\mathbb{R}^{\frac{H}{16}\times\frac{W}{16}\times D\times 256}$.

\begin{figure}[!h]
\begin{center}
\includegraphics[width=\linewidth]{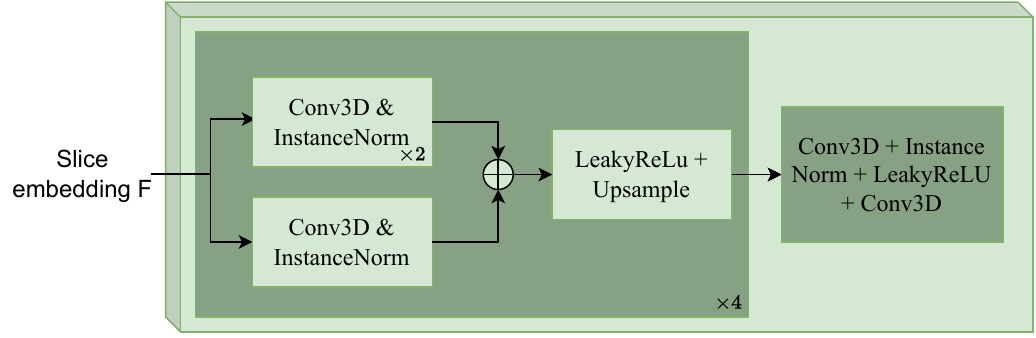}
\vspace{-2em}
\caption{Architecture of the proposed lightweight 3D decoder.} 
\vspace{-2em}
\label{fig:decoder}
\end{center}
\end{figure}

\noindent
\textbf{Decoder. }
Because our decoder must handle 3D volumetric data, we cannot utilize SAM's mask decoder, which is specifically designed for 2D natural images. Instead, we propose the development of an appropriate 3D decoder. However, creating a 3D network with the Vision Transformer \cite{vit2020} and its variants can be resource-intensive, requiring significant computational power and increasing inference time, especially when dealing with a large value of D. Therefore, we suggest the design of a lightweight 3D decoder comprising four 3D convolutional blocks with skip connections \cite{resnet} and a segmentation head, as elaborated in Figure \ref{fig:decoder}.

\begin{table*}[!h]
\setlength{\tabcolsep}{5pt}
\renewcommand{\arraystretch}{0.9}
\centering
\caption{Quantitative results on Synapse dataset.}\label{tab:synapse}
\vspace{-1em}
\resizebox{0.8\linewidth}{!}{
\begin{tabular}{c|c|l|a|bb|dddddddd} 
\hline
\multirow{2}{*}{\textbf{SAM}} & \multirow{2}{*}{\textbf{Networks}} & \multirow{2}{*}{\textbf{Methods}} &  & \multicolumn{2}{c|}{\cellcolor{cambridgeblue}\textbf{Average}} & \multicolumn{8}{c}{\cellcolor{navajowhite}\textbf{DSC on individual abdominal organs}} \\ \hhline{~|~|~|>{\arrayrulecolor{lightcornflowerblue}}->{\arrayrulecolor{black}}----------}
 &  &  & \multirow{-2}{*} {\textbf{Params} $\downarrow$} & HD $\downarrow$ & DSC $\uparrow$ & RKid & LKid & Spl & Gal & Sto & Pan & Aor & Liv\\
 
\hline

 \multirow{10}{*}{\xmark} & \multirow{6}{*}{2D} & TransUNet \cite{transunet} &  96.07M & 31.69 & 77.49 & 77.02 & 81.87 & 85.08 & 63.16 & 75.62 & 55.86 & 87.23 & 94.08\\
 
 & &Swin-Unet \cite{swinunet} & 27.17M & 21.55 & 79.13 & 79.61 & 83.28 & 90.66 & 66.53 & 76.60 & 56.58 & 85.47 & 94.29\\

 & &TransDeepLab \cite{transdeeplab} & 21.14M & 21.25 & 80.16 & 79.88 & 84.08 & 89.00 & 69.16 & 78.40 & 61.19 & 86.04 & 93.53\\

 & &HiFormer-S \cite{hiformer} & 23.25M & 18.85 & 80.29 & 64.84 & 82.39 & 91.03 & 73.29 & 78.07 & 60.84 & 85.63 & 94.22\\
 
 & &HiFormer-B \cite{hiformer} & 25.51M & 14.70 & 80.39 & 79.77 & 85.23 & 90.99 & 65.69 & 81.08 & 59.52 & 86.21 & 94.61\\
 
 & &HiFormer-L \cite{hiformer} & 29.52M & 19.14 & 80.69 & 78.37 & 84.23 & 90.44 & 68.61 & 82.03 & 60.77 & 87.03 & 94.07\\
 
\hhline{~|-|-|-|--|--------}
 
 & \multirow{4}{*}{3D} & MISSFormer \cite{missformer} & - & 18.20 & 81.96  & 82.00 & 85.21 & 91.92 & 68.65 & 80.81 & 65.67 & 86.99 & 94.41\\

 & &nnFormer \cite{nnformer} & 150.50M & 10.63 & 86.57 & 86.25 & 86.57 & 90.51 & 70.17 & 86.83 & 83.35 & 92.04 & 96.84\\
 
 & & UNETR \cite{unetr} & 92.49M & 18.59 & 78.35 & 84.52 & 85.60 & 85.00 & 56.30 & 70.46 & 60.47 & 89.80 & 94.57\\
 
 & &UNETR++ \cite{unetr++} & 42.95M & 7.53 & 87.22 & 87.18 & 87.54 & 95.77 & 71.25 &  86.01 & 81.10 & 92.52 & 96.42\\
\hline
 \multirow{3}{*}{\cmark} & \multirow{2}{*}{2D} & SAMed \cite{samed} & 18.81M & 20.64 & 81.88 & 79.95 & 80.45 & 88.72 & 69.11 & 82.06 & 72.17 & 87.77 & 94.80\\
 
 & &SAMed\_s \cite{samed} & 6.32M & 31.72 & 77.78 & 78.92 & 79.63 & 85.81 & 57.11 & 77.49 & 65.66 & 83.62 & 93.98\\

 \hhline{~|-|-|-|--|--------}
 
 &3D & \textbf{SAM3D (Ours)} & 1.88M & 17.87 & 79.56 & 85.64 & 86.31 & 84.29 & 49.81 & 76.11 & 69.32 & 89.57 & 95.42\\ \hline
\end{tabular}}
\vspace{-1em}
\end{table*}

\noindent
\textbf{Objective Function. }
We train our SAM3D network with a combination loss of both the dice loss and cross-entropy loss. The formulation is as follows:
\vspace{-0.5em}
\begin{equation}
\small
\begin{split}
    \mathcal{L}(Y, \hat{Y}) = - \sum_{n=1}^{N}\sum_{k=1}^{K}( 
    \frac{2 \times Y_{k, n} \hat{Y}_{k, n}} {Y_{k, n} ^ 2 + \hat{Y}_{k, n} ^ 2}
    + Y_{k, n} log \hat{Y}_{k, n}) 
\end{split}
\end{equation}
here, $Y$ is the predicted segmenting result from SAM3D, and $\hat{Y}$ is the ground truth. $N$ represents the number of classes, $K$ denotes the number of voxels, and $Y_{k, n}$ and $\hat{Y}_{k, n}$ refer to the predictions and the ground truths at voxel $j$ for class $i$, respectively.

Additionally, we employ the deep supervision technique for multiple decoding stages. Specifically, the output features of each decoding stage pass through a segmentation block, consisting of one 3 x 3 x 3 and one 1 x 1 x 1 convolution layer, to generate predictions for one typical stage. To calculate the loss value for one typical stage, we down-sample the ground truth to match the prediction resolution. Consequently, the final loss can be defined as follows:
\vspace{-0.2em}
\begin{equation}
    \mathcal{L}_{total} = \sum_{l=1}^{L} \alpha_{l} \times \mathcal{L}_{l}
\end{equation}
here, L is set to 3, representing the number of decoder layers. $\alpha_{l}$ signifies the hyperparameter controlling the contribution of different resolutions to the final loss function. In practice, we set $\alpha_{2} = \frac{\alpha_{1}}{2}$ and $\alpha_{3} = \frac{\alpha_{1}}{4}$ with all $\alpha$ hyperparameters normalized to 1.

\vspace{-1em}
\section{Experiments}
\label{sec:Experiment}
\vspace{-0.5em}
\textbf{A. Datasets and Evaluation Metrics.}

\noindent
\underline{\textit{Datasets:}} We conduct the experiments on four datasets: Multi-organ CT Segmentation (Synapse) \cite{synapse}, Automated Cardiac Diagnosis (ACDC) \cite{acdc}, Brain Tumor Segmentation (BraTS) \cite{lungandbrats}, and Lung Tumor Segmentation (Lung) \cite{lungandbrats}. BraTS and Lung come from the Medical Segmentation Decathlon challenge (MSD) \cite{lungandbrats}. For a fair comparison, we follow the data splitting of previous works, e.g. nnFormer \cite{nnformer} and UNETR++ \cite{unetr++}. 

\begin{table}[!htb]
\vspace{-1em}
\setlength{\tabcolsep}{5pt}
\renewcommand{\arraystretch}{1.0}
\centering
\caption{Quantitative results on ACDC dataset.}\label{tab:acdc}
\vspace{-1em}
\resizebox{0.9\linewidth}{!}{
\begin{tabular}{l|a|b|ddd} 
\hline
  \multirow{2}{*}{\textbf{Methods}} &  & {\textbf{Average}} & \multicolumn{3}{d}{\cellcolor{navajowhite}\textbf{DSC on individual regions}}\\ \hhline{~|>{\arrayrulecolor{lightcornflowerblue}}->{\arrayrulecolor{black}}----}
 &  \multirow{-2}{*} {\textbf{Params}$\downarrow$} & DSC $\uparrow$ & \hspace*{2mm} RV \hspace*{2mm} & \hspace*{3mm} LV \hspace*{3mm} & MYO \\
\hline
 TransUNet \cite{transunet} & 96.07M & 89.71 & 88.86 & 84.54 & 95.73\\
 Swin-Unet \cite{swinunet} & 27.17M & 90.00 & 88.55 & 85.62 & 95.83\\
 UNETR \cite{unetr} & 92.49M & 86.61 & 85.29 & 86.52 & 94.02\\
 MISSFormer \cite{missformer} & -- & 87.90 & 86.36 & 85.75 & 91.59\\
 nnFormer \cite{nnformer} & 150.5M & 92.06 & 90.94 & 89.58 & 95.65\\
 UNETR++ \cite{unetr++}& 66.80M & 92.83 & 91.89 & 90.61 & 96.00\\
\hline
 \textbf{SAM3D (Ours)} & 1.88M & 90.41 & 89.44 & 87.12 & 94.67\\
\hline
\end{tabular}}
\end{table}

\begin{table}[!htb]
\centering
\caption{Quantitative results on Lung dataset.} 
\vspace{-1em}
\setlength{\tabcolsep}{15pt}
\renewcommand{\arraystretch}{1.0}
\label{tab:lung}
\resizebox{0.8\linewidth}{!}{
\begin{tabular}{l|a|b} 
\hline
 \textbf{Methods} & \textbf{Params} $\downarrow$ & \textbf{Average DSC} $\uparrow$\\
\hline
 nnUNet \cite{nnunet} & -- & 74.31\\
 Swin UNETR \cite{swinunetr} & 62.83M & 75.55\\
 nnFormer \cite{nnformer} & 150.5M & 77.95\\
 UNETR \cite{unetr} & 92.49M & 73.29\\
 UNETR++ \cite{unetr++} & 121.17M & 80.68\\
\hline
 \textbf{SAM3D (Ours)} & 1.88M & 71.42\\
\hline
\end{tabular}}
\vspace{-1em}
\end{table}

\begin{table}[!htb]
\centering
\setlength{\tabcolsep}{2pt}
\renewcommand{\arraystretch}{1.0}
\caption{Quantitative results on BraTS dataset.} 
\vspace{-1em}
\label{tab:brats}
\resizebox{\linewidth}{!}{
\begin{tabular}{l|a|bb|dd|dd|dd} 
\hline
 \multirow{2}{*}{\textbf{Methods}} &  & \multicolumn{2}{b|}{\cellcolor{cambridgeblue}\textbf{Average}} & \multicolumn{2}{d|}{\cellcolor{navajowhite}\textbf{WT}} & \multicolumn{2}{d|}{\cellcolor{navajowhite}\textbf{ET}} & \multicolumn{2}{d}{\cellcolor{navajowhite}\textbf{TC}}\\ \hhline{~|>{\arrayrulecolor{lightcornflowerblue}}->{\arrayrulecolor{black}}--------}

 & \multirow{-2}{*} {\textbf{Params}$\downarrow$} & HD $\downarrow$ & DSC $\uparrow$ & HD $\downarrow$ & DSC $\uparrow$ & HD $\downarrow$ & DSC $\uparrow$ & HD $\downarrow$ & DSC $\uparrow$ \\
\hline
TransUNet \cite{transunet} & 96.07M & 12.98 & 64.4 & 14.03 & 70.6 & 10.42 & 54.2 & 14.50 & 68.4\\
UNETR \cite{unetr} & 92.49M & 8.82 & 71.1 & 8.27 & 78.9 & 9.35 & 58.5 & 8.85 & 76.1 \\
nnFormer \cite{nnformer} & 150.5M & 4.05 & 86.4 & 3.80 & 91.3 & 3.87 & 81.8 & 4.49 & 86.0\\
UNETR++ \cite{unetr++} & 42.65M & 5.85 & 77.7 & 4.79 & 91.2 & 4.22 & 78.5 & 6.78 & 78.4 \\
\hline
 \textbf{SAM3D (Ours)} & 4.63M & 8.72 & 72.9 & 6.03 & 88.0 & 10.05 & 69.6 & 9.79 & 76.6 \\
\hline
\end{tabular}}
\vspace{-1.2em}
\end{table}


\begin{table*}[!t]
\centering
    \vspace{-1em}
    \setlength{\tabcolsep}{2pt}
    \centering 
    \caption{Ablation study of the skip connection in our lightweight 3D decoder on ACDC and Synapse datasets.}\label{tab:ablation}
    \vspace{-1em}
    \resizebox{\linewidth}{!}{
    \setlength{\tabcolsep}{2.8pt}
    \subfloat[ACDC dataset. \vspace{-0.5em}]{
    \resizebox{0.4\linewidth}{!}{
 \begin{tabular}{l|b|ddd} 
\hline
  \multirow{2}{*}{\textbf{Settings}} & {\cellcolor{cambridgeblue}\textbf{Average}}  & \multicolumn{3}{d}{\cellcolor{navajowhite}\textbf{DSC on individual regions}} \\ \hhline{~|-|---}
  & \textbf{DSC} $\uparrow$ & \hspace*{3mm} RV \hspace*{3mm} & \hspace*{3mm} LV \hspace*{3mm} & MYO \\ 
\hline
 w/o skip connection & 89.73 & 88.46 & 94.41 & 86.32 \\
\hline
 w skip connection & 90.41  & 89.44 & 94.67 & 87.12\\
\hline
\end{tabular}}}
    \quad
    \subfloat[Synapse dataset. \vspace{-0.5em}]{
    \resizebox{0.6\linewidth}{!}{
 \begin{tabular}{l|bb|dddddddd} 
\hline
 \multirow{2}{*}{\textbf{Settings}} & \multicolumn{2}{b|}{\cellcolor{cambridgeblue}\textbf{Average}}  & \multicolumn{8}{d}{\cellcolor{navajowhite}DSC\textbf{ on individual abdominal organs}}\\ \hhline{~|--|--------}
 & \textbf{HD}$\downarrow$ & \textbf{DSC}$\uparrow$ & RKid & LKid & Spl & Gal & Sto & Pan & Aor & Liv\\
\hline
 w/o skip connection & 25.87 & 79.33 & 84.68 & 85.20 & 85.26 & 50.55 & 75.07 & 68.83 & 90.10 & 94.98\\
\hline
 w skip connection & 17.87 & 79.56 & 85.64 & 86.31 & 84.29 & 49.81 & 76.11 & 69.32 & 89.57 & 95.42\\
\hline
\end{tabular}}}
}
\vspace{-2em}
\label{abla:hyper-param}
\end{table*}



\vspace{-2mm}
\noindent
\underline{\textit{Metrics:}} We evaluate the network's accuracy using the Dice Similarity Coefficient (DSC) and the 95\% Hausdorff Distance (HD95), while the network's complexity is measured by the number of trainable parameters (\#params). The HD95 is calculated based on the 95th percentile of the distances between the boundaries of predictions and ground truths.

\noindent
\textbf{B. Implementation Details.}

\noindent Our model is implemented based on Python 3.8.10 with PyTorch library and trained on a single NVIDIA RTX 2080 Ti GPU with 11GB memory. We use \textbf{ViT-B} version as our backbone for the SAM's image encoder due to the limited resources. Instead of exhaustively finding an overfitting training procedure, we trained our model with the general training strategy of nnFormer \cite{nnformer} and UNETR++ \cite{unetr++}, the stochastic gradient descent (SGD) with a momentum of 0.99 and a weight decay of 3e-5. The learning rate scheduler is defined as $lr = init\_lr \times (1 {}-{} \frac{epoch}{max\_epoch})^{power}$, where $init\_lr$ = 1e-2, $power$ = 0.9, and $max\_epoch$ = 1000. One epoch consists of 250 iterations. For ACDC, Synapse, BraTS, and Lung datasets, SAM3D is trained with the 3D volume sizes of 160 x 160 x 14, 176 x 176 x 64, 64 x 64 x 64 and 192 x 192 x 34, respectively. We also utilize the same data augmentation techniques including rotation, scaling, brightness adjustment, gamma augmentation, and mirroring. The batch size is set to 4 for ACDC and 2 for Synapse, BraTS, and Lung.

\noindent
\textbf{C. Performance Comparisons.}

\noindent
We compared our SAM3D with recent SOTA methods on both CNNs-based networks, e.g. nnFormer \cite{nnformer} and Transformer-based networks, e.g. TransUNet \cite{transunet}, Swin-Unet \cite{swinunet}, TransDeepLab \cite{transdeeplab}, HiFormer \cite{hiformer}, MISSFormer \cite{missformer}, UNETR \cite{unetr} and SAM-based models SAMed and SAMed\_s \cite{samed}. The performance comparisons are reported in Tables \ref{tab:synapse}, \ref{tab:acdc}, \ref{tab:lung}, and \ref{tab:brats} including both accuracy (i.e. HD95 and DSC metrics) and network complexity (\#params). 

Synapse comprises eight abdominal organs in a large dataset and the performance comparison is shown in Table \ref{tab:synapse}. Among the models evaluated, UNETR++ (a Transformer-based model) achieved the best results with 42.9M parameters, while nnFormer ranked second with 150.5M parameters. Notably, SAMed\_s distinguishes itself by achieving impressive results with a modest 6.32M parameters and a DSC of 77.78\%. SAMed\_s shares a similar architecture with our SAM3D, fine-tuned from SAM, but differs in processing methods. SAMed\_s employs a straightforward slice-by-slice approach, while SAM3D considers depth-wise information. Despite this difference, both models are efficient in parameter usage. SAMed\_s requires 6.32M parameters, whereas SAM3D excels with just 1.88M parameters. Furthermore, SAM3D achieves a DSC score exceeding 1.78\%, demonstrating superior performance compared to SAM-based methods with lightweight models.

While SAMed is exclusive to the Synapse dataset, our SAM3D can be evaluated on a variety of other datasets, including Cardiac, Brain Tumor, and Lung. In Table \ref{tab:acdc} and \ref{tab:lung}, it is evident that SAM3D competes favorably with SOTA  CNNs/Transformer-based networks on the Cardiac ACDC and Lung datasets. For instance, SAM3D surpasses TransUnet's performance on the ACDC dataset with a 0.41\% increase in DSC while utilizing less than 50$\times$ the number of parameters. Table \ref{tab:brats} further illustrates SAM3D's competitiveness with other leading models on the Brain Tumor Brats dataset, despite its significantly lower parameter count. For example, SAM3D achieves a 1.8\% DSC improvement compared to UNETR, while requiring less than 20$\times$ the number of params. It is worth noting that the MRI scans in Brats contain four modalities, which explains SAM3D's parameter count being four times that of other single-modality models.

\begin{figure}[!h]
\begin{center}
\includegraphics[width=\linewidth]{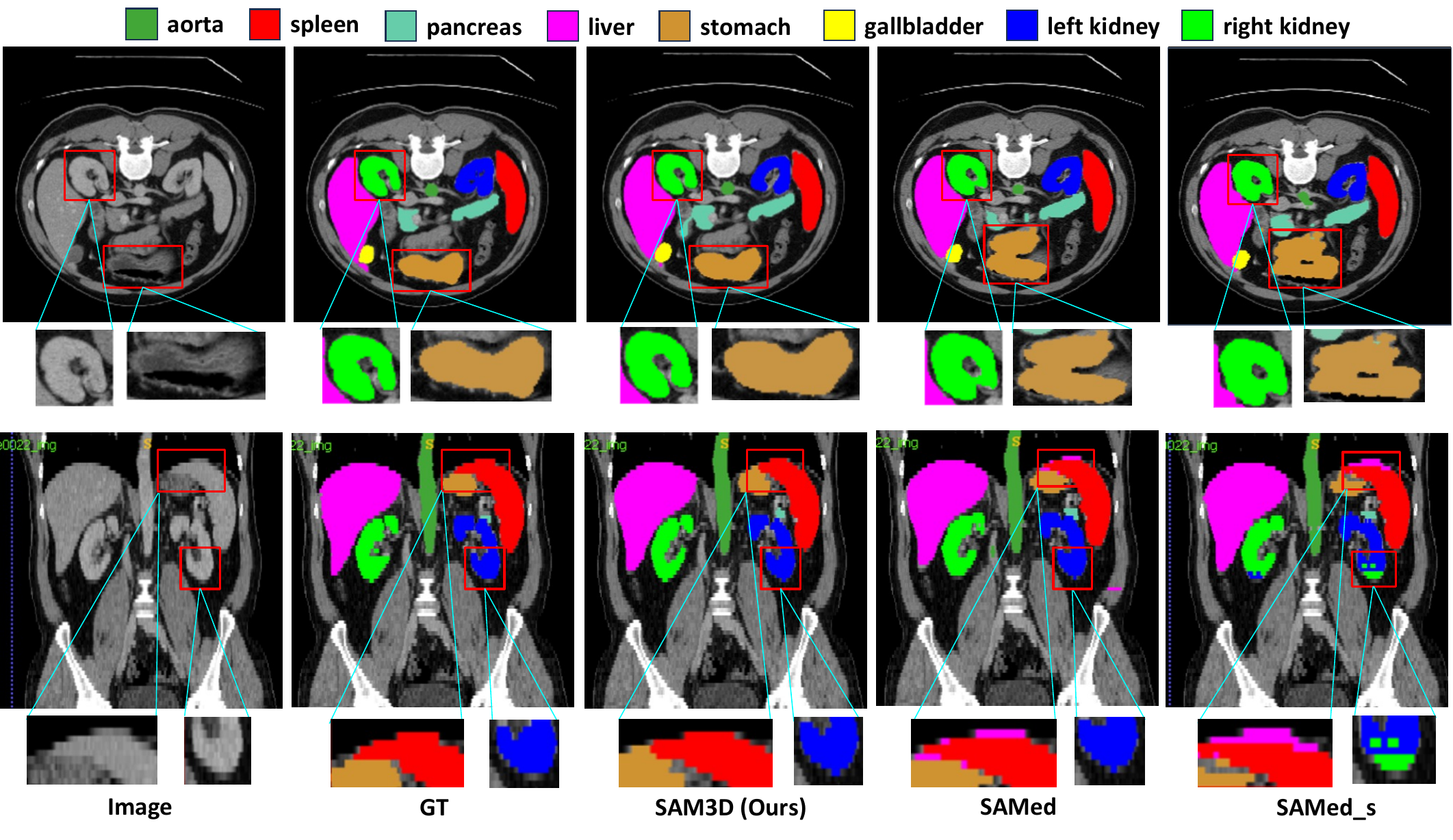}
\vspace{-2em}
\caption{Qualitative comparison between our SAM3D ($3^{rd}$ column) and other SAM-based volumetric segmentation models SAMed ($4^{th}$ column) and SAMed\_s ($5^{th}$ column) on Synapse dataset. SAMed and SAMed\_s require 18.81M and 6.32M params whereas our SAM3D needs only 1.88M.} 
\vspace{-2em}
\label{fig:qualitative}
\end{center}
\end{figure}

Fig. \ref{fig:qualitative} visually presents samples from the Synapse dataset. In this illustration, we compare our approach (in the third column) with the outcomes obtained from SAMed and SAMed\_s, which represent SOTA in SAM-based methods for volumetric medical image segmentation. Despite the reduced trainable params, SAM3D exhibits superior segmentation performance compared to the other two methods.

\noindent 
\textbf{D. Ablation Study.}

\noindent To assess the impact of skip connections in our proposed lightweight 3D decoder, we conducted an ablation study on ACDC and Synapse datasets as depicted in Table \ref{tab:ablation}. The results clearly indicate that these skip connections contribute positively to the model's performance, resulting in improvement. We believe that these skip connections play a crucial role in preserving information related to edges and boundaries from lower-level features, enhancing the precision of the segmentation process.


\vspace{-1em}
\section{Conclusion}
\label{sec:Conclusion}
\vspace{-0.5em}
In this study, we introduce SAM3D, an efficient and simple SAM-based model tailored for volumetric medical image segmentation. Our approach harnesses the capabilities of a SAM pre-trained encoder coupled with a lightweight 3D decoder. Through extensive experimentation, we have established that SAM3D competes effectively with current SOTA 3D neural networks and Transformer-based models while demanding significantly fewer parameters (50$\times$ fewer). Furthermore, SAM3D outperforms other lightweight networks in the context of volumetric segmentation. As SAM has already made a substantial impact on natural image segmentation, our research extends its potential to the domain of medical image segmentation. We anticipate that this work will serve as an inspiration for future researchers, fostering advancements in the field of medical segmentation

\noindent \textbf{Discussion.} In our experiments, we employed the smallest SAM variant, which utilizes ViT-B backbone, primarily due to resource and time constraints. We hypothesize that ViT-L and ViT-H pre-trained models may yield even more remarkable results. Consequently, we encourage researchers to explore these options for our segmentation task.

Additionally, our simple decoder leaves room for developing a more complex architecture, which could potentially enhance the model's performance. This presents a promising avenue for further research and development.



\section{Acknowledgement}

Nhat-Tan Bui and Ngan Le are supported by the National Science Foundation (NSF) under Award No OIA-1946391 RII Track-1, NSF 1920920 RII Track 2 FEC, NSF 2223793 EFRI BRAID, NSF 2119691 AI SUSTEIN, NSF 2236302.
Minh-Triet Tran is sponsored by Vietnam National University Ho Chi Minh City (VNU-HCM) under grant number DS2020-42-01.
Dinh-Hieu Hoang is funded by Vingroup Joint Stock Company and supported by the Domestic Master/ PhD Scholarship Programme of Vingroup Innovation Foundation (VINIF), Vingroup Big Data Institute (VINBIGDATA), code VINIF.2022.ThS.JVN.04.

\small
\bibliographystyle{IEEEbib} 
\bibliography{strings}

\end{document}